\documentstyle[twoside,fleqn,espcrc2,epsfig]{article}

\newcommand{\be}{\begin{eqnarray}}
\newcommand{\ee}{\end{eqnarray}}
\newcommand{\ben}{\begin{eqnarray*}}
\newcommand{\een}{\end{eqnarray*}}
\newcommand{\la}{\langle}
\newcommand{\ra}{\rangle}

\def\lab#1      {\hbox{\small #1} }
\def\diffn#1	  {\Delta^{-}_{#1}}
\def\mb#1       {\mbox{\boldmath $#1$}}
\newcommand{\AmS}{{\protect\the\textfont2
  A\kern-.1667em\lower.5ex\hbox{M}\kern-.125emS}}

\title{Vortices and monopole distributions in $Z(2) \times SO(3)$ lattice gauge theory}
\author{Andrei Alexandru and Richard W. Haymaker\thanks{Presented 
        by R. Haymaker at Lattice 2000, July 17 - 22, Bangalore.
   Work partially supported by the U. S. Department of Energy under grant 
               DE-FG05-91 ER 40617.}
               \address{Dept. of Physics and Astronomy, 
       	Louisiana State University, Baton Rouge, Louisiana,
       	70803  USA}        }

\begin{document}

\begin{abstract}
We examine the occurance of $Z(2)$ and $SO(3)$ vorticies and monopole distributions in the neighborhood of
Wilson loops. 
We use the Tomboulis formulation, equivalent to the Wilson action, in which the links are invariant under $Z(2)$ 
transformations and new plaquette variables carry the $Z(2)$ degrees of freedom. This gives new gauge 
invariant observables to help gain insight into the area law and  structure of the flux tube.
\end{abstract}

% typeset front matter (including abstract)
\maketitle

%\section{Introduction}
$SU(N)$ lattice gauge theory with a Wilson action can be reformulated in terms of 
$Z(N)$ and $SU(N)/Z(N)$ variables as derived by Tomboulis\cite{t} and 
Kovacs and Tomboulis\cite{kt}.  We report results of simulations in
these variables using an algorithm described elsewhere\cite{ah}.  

For the case considered here 
the $SU(2)$ group summation becomes an $SU(2)/Z(2) = SO(3)$ integration over the links (bonds), $U(b)$,
and a discrete sum over the independent Z(2) variables, $\{\sigma(p)\}$, living on plaquettes.  
There are also dependent plaquette $Z(2)$ variables, $\{\eta(p)\}$,  functions of $\{U(b)\}$,
defined by $Tr[U(\partial p)] = \eta(p)|Tr[U(\partial p)]|$.  
\ben
Z  
=
\int 
\left[
     d U(b)
\right]
\sum_{\sigma(p)}
\prod_c \delta[\sigma(\partial c)\eta(\partial c)] \times
\\
\exp
\left(
   \beta 
   \sum_{p} 
   \frac{1}{2} |\lab{tr} [  U(\partial p)]| 
   \sigma(p)
\right).
\label{z36}
\een
The expression for a Wilson loop, $W_C$, 
includes a tiling of any surface $ S = \partial C $,
\be
W_C = \sigma(S) \eta(S) \frac{1}{2} tr[U(\partial S)].
\ee
Note that $W_{1 \times 1} = \sigma(p) \frac{1}{2}|tr[U(\partial p)]|$.

The excitations in this formulation 
include co-closed $2d$ vortex sheets which provide a mechanism to disorder the Wilson 
loop. They are more easily visualized in the dual representation where 
they consist of closed tiled  sheets of either 
negative $\sigma(p)$ or negative $\eta(p)$ variables living on dual plaquettes.  
Each species form `open vortex patches', (which we
call `patches')
on the surface bounded by its corresponding species of a closed monopole loop living on dual links.
We denote the boundary of patches of $\sigma(p) = -1$ as a $Z(2)$ monopole
current and similarly $SO(3)$ monopole current surrounding the $\eta(p) = -1$ patches. 

Constraints in the partition function enforce this vortex structure by requiring that any 
$Z(2)$ monopole loop be coincident with an $SO(3)$ monopole loop thus closing the surface. (This   
is the dual description of the cubic constraints in $Z$.)
This gives  a `hybrid' vortex.  The degenerate cases consist of a 
pure $\sigma(p)$ or a pure $\eta(p)$ vortex. 

We are interested in sign fluctuations which disorder the Wilson loop. 
In order to clarify the simulation results below, consider first a simplified 
configuration $\{U(b),\sigma(p)\}$ for 
which a particular  Wilson loop, has the value $ = -1$ and further all links on $C$ $= I$, and only one 
of the tiling factors in Eqn.(1) is $-1$.   And we also take a particular 
spanning surface $S$  e.g. the minimal surface.  

\begin{enumerate}

\item Suppose that all $\sigma(p)= \eta(p) = +1$ 
on $S$ except for one negative $\sigma(p)$.  

\item Then we can conclude that
{\em either} (i) a $\sigma(p)$ vortex links the loop {\em or} 
(ii) a hybrid vortex links the loop with a $\sigma(p) = -1$ 
patch occurring on this particular surface.  

\item Consider all distortions of $S$.  If the negative 
sign is found to switch from a $\sigma(p)$ 
to the $\eta(p)$, then this is case (ii), a hybrid vortex links the loop.  

\item If the signs of $\eta(p)$ and $\sigma(p)$ do not depend on $S$ then we are seeing case (i),
a $\sigma(p)$ vortex linking the loop.

\item Suppose instead all $\sigma(p)= \eta(p) = +1$ on $S$ except for one negative $\eta(p)$ (instead of 
one negative $\sigma(p)$), and that this persists for all distortions of $S$ then we are seeing case(iii), an $\eta(p)$ vortex linking the loop.

\end{enumerate} 

The $\sigma(p)$ vortices (or patches) are known as `thin' vortices (or patches). Thin vortices are suppressed
at large $\beta$ because they cost action proportional to the vortex area and can at most disorder the 
perimeter of a Wilson loop.  However thin patches do not suffer this limitation and indeed do contribute
to Wilson loops in the data reported here.

The $\eta(p)$ vortices (or patches) are indicators of true `thick' vortices (or patches) due to vorticity
in  $\{U(b)\}$.   This is complicated by the fact $\eta(p)$ patches can be distorted with no 
cost of action because the links are SO(3) configurations, invariant under flipping the signs
of links.  An $\eta(p)$ vortex or patch can be moved to change the linkage  number in $C$.  However in
the above example this would flip the sign of a link in $C$.  
The combination   $\eta(S) \frac{1}{2} tr[U(\partial S)]$ is invariant under these sign
flips and we use this to detect the presence of a thick vortex patch piercing $S$.   

Following the studies in related work by Greensite et. al. \cite{g} on projection vortices 
we use linkage numbers to tag Wilson loops and segregate then before computing averages.
We count patches, mod 2, {\em piercing the minimal surface} using the operators\cite{t,kt}
\be
{\cal N}_{\lab{thin patch} } &=& \sigma_S,  \\
{\cal N}_{\lab{thick patch} } &=& \eta_S  \lab{sgn} \{tr[U(\partial S)]\}.
\ee
(Since we do not measure on every $S$, we can not discriminate between  hybrid and thin
or hybrid and thick linkage numbers.)

%\section{Numerical results}
Fig. 1 gives the fraction  $X_e$ of Wilson loops tagged to have  0 mod 2 vortices as a function
of Wilson loop area on a $12^4$ lattice at $\beta = 2.5$ ($X_e + X_o = 1$, where $X_o$ is the 1 mod 2 fraction).
The dashed curve corresponds to ${\cal N}_{\lab{thin patch} }$, 
the dotted curve to ${\cal N}_{\lab{thick patch} }$ and the solid curve to the product, i.e. tagged by the
sign of the Wilson loop itself.
\begin{figure}[h]
\epsfig{file=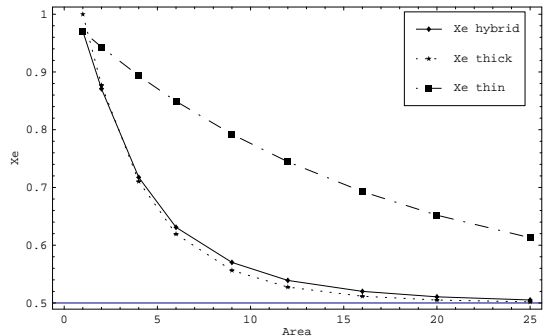,width=7.5cm}
\caption{$X_e$ vs. area}
\end{figure}
For large areas, all curves approach $0.5$ giving nearly equal probabilites of an even or odd vortex number.
Qualitatively, the rate of approach is a measure of the number of vortices per unit area 
piercing the minimal surface $S$. Clearly the thin patches are the least dense in this sense.  

An interesting feature is that two curves cross.  If the occurance of thin and thick patches were statistically
independent, then counting either one (solid line), would be closer to the asymptotic value of  $X_e$ and hence
must lie below the two individual cases.  A non-zero probablity of pairing of thin and thick patches might
account for this crossing.
\begin{figure}[h]
\epsfig{file=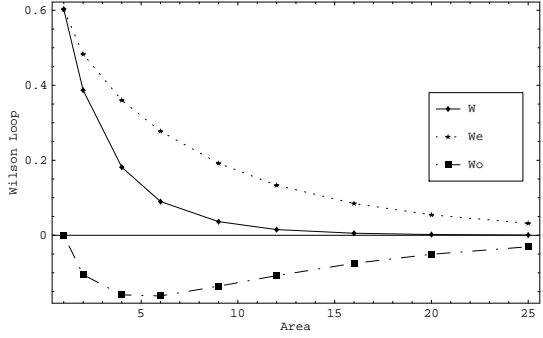,width=7.5cm}
\caption{$W,W_e,W_o$ vs. area for $\beta = 2.3$}
\end{figure}
\begin{figure}[h]
\epsfig{file=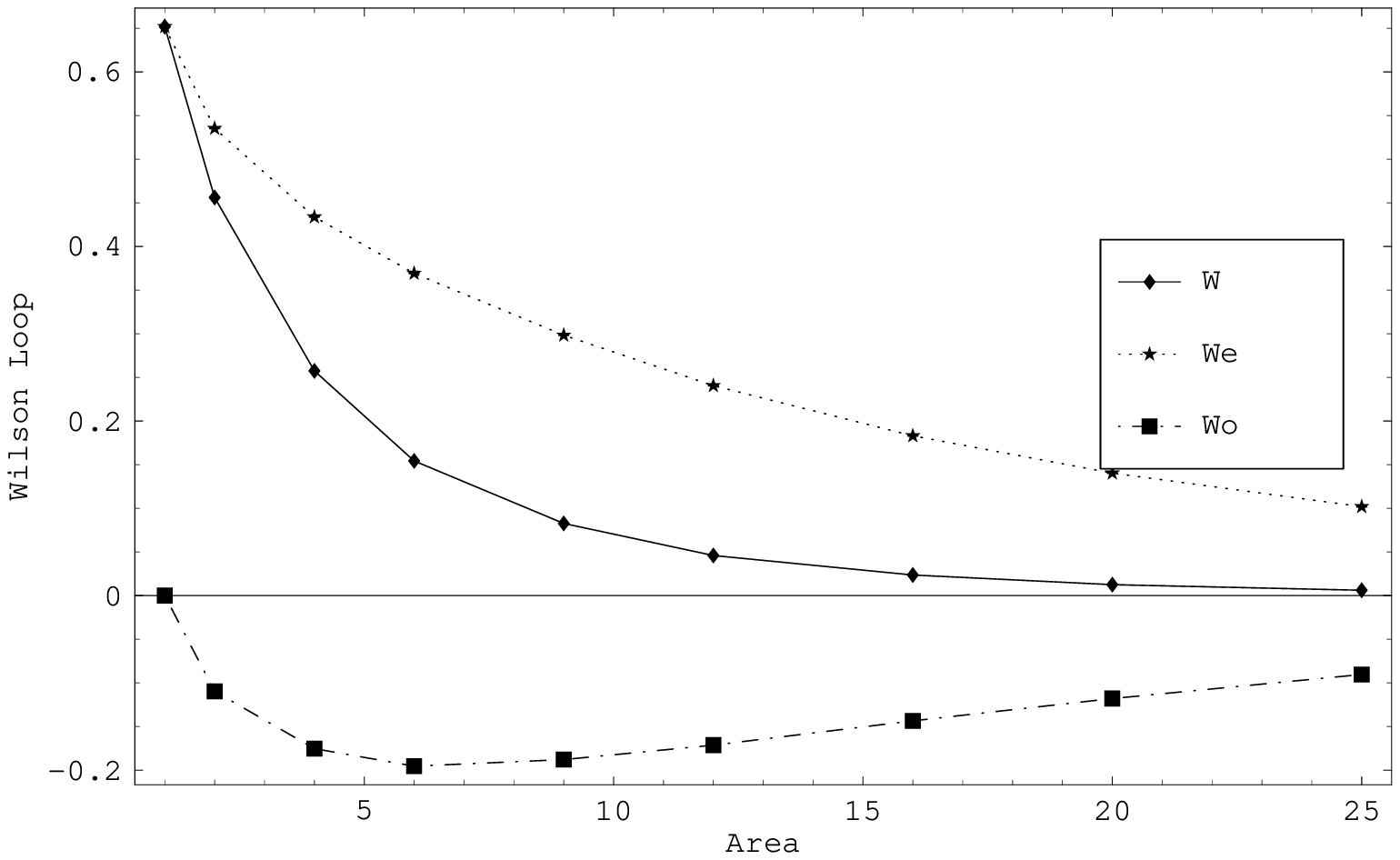,width=7.5cm}
\caption{$W,W_e,W_o$ vs. area for $\beta = 2.5$}
\end{figure}
\begin{figure}[h]
\epsfig{file=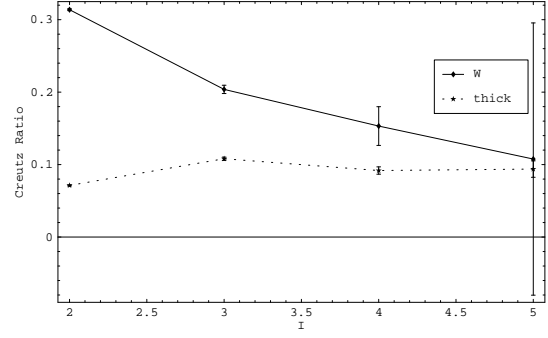,width=7.5cm}
\epsfig{file=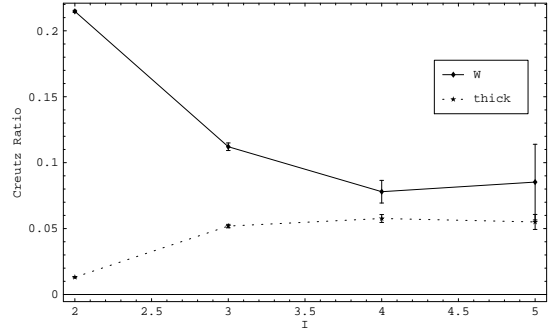,width=7.5cm}
\caption{creutz raios $\chi(I,I)$ (solid line)  and $\chi_e(I,I)$ (dotted line)
for $\beta =$ $2.3$ (upper) and $2.5$ lower.}
\end{figure}

Figs. 2 and 3 give  $W, W_e$ and $W_o$ as a function of area for $\beta = 2.3$ and 
$\beta = 2.5$.  The dotted curve, $W_e$ corresponds to ${\cal N}_{\lab{thick patch} } =  0 $, 
the dashed curve, $W_o$ to ${\cal N}_{\lab{thick patch} } = 1 $ and the solid curve, 
$W$ to the Wilson loop itself. 
The values of $W_e$ and $W_o$ at area $=1$ follows from Eqns.(1) and (3). The exponential fall off follows from the
fact that thin patches are still active in disordering this loop.  

The dotted lines in Fig.4  are Creutz ratios, 
$\chi_e (I,I)=- \ln \frac{W_e(I,I) W_e(I-1,I-1)}{W_e(I,I-1)^2}$.  Hence $W_e$ is showing an area 
law due to thin patches alone disordering the loop.   We also plot $\chi (I,I) $
corresponding to $W$ for comparison.  Poor statistics precludes a scaling analysis. Nevertheless
the disordering due to thin patches compared to the full disordering is very similar for this range of
$\beta$.

Finally we report the monopole density, $j_m$
\ben
\beta = 2.3:  &&  0.2156(4)\\
\beta = 2.5:  &&   0.142(1)
\een
We also measured this within the flux tube
\ben
\la j_m \ra_W = \la W j_m \ra/\la W \ra - \la j_m \ra.
\een
We found that the monopole density was suppressed there. Details will appear elsewhere.

We thank E.T.Tomboulis for helpful discussions.

\end{document}